\documentclass[prr,reprint,numerical]{revtex4-1}
\usepackage{graphicx}  
\usepackage{dcolumn}   
\usepackage{bm}        
\usepackage{amssymb}   
\usepackage{amsmath}

\usepackage[T1]{fontenc}
\usepackage[latin9]{inputenc}
\usepackage{amsthm}
\usepackage{xcolor}

\usepackage{tikz}

\graphicspath{{./img/}}

\newcommand{\gratio}{\gamma^{\rm r}}
\newcommand{\kr}{k^{\rm r}}
\newcommand{\meanN}{\langle N\rangle}
\newcommand{\der}{\mathrm{d}}
\newcommand{\avg}[1]{\langle #1\rangle}
\newcommand{\Nt}{\mathrm{N}^{\rm t}}
\newcommand{\fc}{f^{\rm c}}
\newcommand{\fd}{f^{\rm d}}
\newcommand{\fm}{f^{\rm c,m}_{1,2}}
\newcommand{\Nc}{\mathrm{N}^{\rm c}}
\newcommand{\kzero}{\kappa^{0}}

\makeatletter
\DeclareOldFontCommand{\rm}{\normalfont\rmfamily}{\mathrm}
\DeclareOldFontCommand{\sf}{\normalfont\sffamily}{\mathsf}
\DeclareOldFontCommand{\tt}{\normalfont\ttfamily}{\mathtt}
\DeclareOldFontCommand{\bf}{\normalfont\bfseries}{\mathbf}
\DeclareOldFontCommand{\it}{\normalfont\itshape}{\mathit}
\DeclareOldFontCommand{\sl}{\normalfont\slshape}{\@nomath\sl}
\DeclareOldFontCommand{\sc}{\normalfont\scshape}{\@nomath\sc}
\makeatother

\usepackage{hyperref}
\usepackage{mathtools}
\usepackage{epstopdf}

\begin{document}
\title{Stability of heterogeneous parallel-bond adhesion clusters under static load}
\author{Anil K. Dasanna}
\author{Gerhard Gompper}
\author{Dmitry A. Fedosov}
\email{d.fedosov@fz-juelich.de}
\affiliation{
Theoretical Physics of Living Matter, Institute of Biological Information Processing and Institute for Advanced Simulation, Forschungszentrum J\"ulich, 52425 J\"ulich, Germany
}	
	
\begin{abstract}
Adhesion interactions mediated by multiple bond types are relevant for many biological and soft matter systems,
including the adhesion of biological cells and functionalized colloidal particles to various substrates. To elucidate 
advantages and disadvantages of multiple bond populations for the stability of heterogeneous adhesion clusters of 
receptor-ligand pairs, a theoretical model for a homogeneous parallel adhesion bond cluster under constant loading is extended to 
several bond types. The stability of the entire cluster can be tuned by changing densities of different bond populations 
as well as their extensional rigidity and binding properties. In particular, bond extensional rigidities determine the 
distribution of total load to be shared between different sub-populations. Under a gradual increase of the total load, the 
rupture of a heterogeneous adhesion cluster can be thought of as a multistep discrete process, in which 
one of the bond sub-populations ruptures first, followed by similar rupture steps of other sub-populations 
or by immediate detachment of the remaining cluster. This rupture behavior is qualitatively independent of involved bond 
types, such as slip and catch bonds. Interestingly, an optimal stability is generally achieved when the total cluster load 
is shared such that loads on distinct bond populations are equal to their individual critical rupture forces. 
We also show that cluster heterogeneity can drastically affect cluster lifetime.   	
\end{abstract}
\maketitle

\section{Introduction}

Adhesion interactions via receptor-ligand bonds are essential for many biological and soft matter systems. Examples include 
cell adhesion \cite{Gumbiner_CAM_1996,Discher_TCF_2005,Schwarz_UWS_2012,Sackmann_PCA_2014} and migration 
\cite{Paluch_FAM_2016,Collins_RWN_2015}, synapse formation \cite{Grakoui_TIS_1999,Qi_SPF_2001}, 
adhesion of lipid vesicles \cite{Cuvelier_HDV_2004,Smith_FIG_2008,Murrell_LAS_2014} and drug-delivery carriers 
\cite{Decuzzi_ASP_2006,Cooley_IPS_2018} to a substrate. Such adhesive interactions depend on the properties of 
receptors and ligands (e.g. density, kinetic rates, mobility) and the characteristics of adhered particles (e.g. size, shape, 
deformability). For instance, binding/dissociation rates of receptors and their mobility together with membrane constraints strongly 
affect the formation of immunological synapse characterized by a highly organized pattern of receptor proteins 
\cite{Qi_SPF_2001,Weikl_PFA_2004}.  In addition to bond rates and mobility, membrane/substrate deformation and 
applied stresses play an important role in the nucleation of bond domains \cite{Bihr_NLR_2012} and their growth and 
distribution \cite{Smith_FIG_2008,Jiang_ADB_2015,Brochard_AIB_2002,Weikl_AMC_2009}.              
 
Bond-mediated adhesion interactions often involve more than one type of receptor-ligand pairs with distinct intrinsic properties. For instance, leukocytes before 
extravasation first bind to and roll at an endothelial cell layer, then show a firm adhesion at the surface
\cite{Ley_LIV_1995,Ebnet_MML_1999,Sorokin_IEM_2010}. This process is facilitated by the ability of P-selectin glycoprotein (PSGL-1) at 
the surface of leukocytes to bind to both selectin and integrin molecules expressed at endothelial cells. Another example 
is the adhesion of malaria-infected red blood cells to the endothelium, in order to avoid their removal in the  spleen
\cite{Berendt_SPF_1990,Udeinya_FMI_1981,Miller_PBM_2002}. Here, Plasmodium falciparum erythrocyte membrane receptor (PfEMP-1)
can bind to multiple ligands (e.g. CD36, ICAM-1, and CSA molecules) at the surface of endothelial cells
\cite{Helms_MCM_2016,Lansche_SCT_2018,Lim_SMC_2017}. Even though it is hypothesized that they act synergistically 
\cite{Yipp_SMA_2000}, the exact roles of different receptor-ligand pairs remain largely unknown. 

Biological cells often interact with a substrate through a number of localized adhesion sites called focal adhesions \cite{Schwarz_UWS_2012}, 
which can be thought of as localized clusters of adhesive bonds under applied stress. Similarly, leukocyte adhesion can be approximated
through adhesive interactions of several discrete clusters, since PSGL-1 proteins are primarily located at the cell's microvilli tips \cite{Bruehl_QLS_1996}. 
Furthermore, PfEMP-1 receptors are positioned at cytoadherent knobs, representing discrete adhesion clusters on the surface of malaria-infected 
erythrocytes \cite{Watermeyer_SSU_2015}. The first simple theoretical model for local adhesion clusters with a single bond population 
was proposed by Bell \cite{Bell_MSA_1978}, who used a mean-field approach to describe the stability of a parallel adhesion-bond cluster of fixed size under constant loading. 
Later, the original model has been extended to dynamic loading \cite{Seifert_RPB_2000,Li_VSM_2016} and generalized to a stochastic model 
for parallel bond clusters \cite{Erdmann_SAC_2004,Erdmann_SDA_2004,Schwarz_FAC_2013}, which shows that the cluster lifetime is always finite
and increases exponentially with the number of bonds within the cluster. Note that these models are applicable to local adhesion clusters with 
a fixed number of adhesive sites (i.e. without receptor mobility).  Furthermore, they can be used to quantify the adhesion of functionalized 
rigid particles used in self-assembled functional materials \cite{Wang_DNA_2015,Zhang_DNA_2017} or for drug delivery \cite{Decuzzi_ASP_2006,Cooley_IPS_2018}.  

In the theoretical model for a homogeneous adhesion bond cluster, each bond can form with a constant on-rate $\kappa^{\rm on}$ and rupture with an off-rate 
$\kappa^{\rm off}(F)$ which depends on the applied force $F$. Note that the ratio $\kappa^{\rm on}/\kappa^{\rm off}$ represents a Boltzmann factor related to the energy change 
due to bond formation, and therefore, it characterizes binding strength. The original model by Bell \cite{Bell_MSA_1978} 
considered a so-called {\it slip} bond with $\kappa^{\rm off} = \kzero \exp{(F/ \fd)}$, where $\kzero$ is the unstressed off-rate and 
and $\fd$ is a characteristic force scale (typically a few $\textrm{pN}$), such that the bond lifetime decreases with increasing $F$ \cite{Evans_PRB_2001}. 
Some biological bonds may behave differently, so-called {\it catch} bonds, such that their lifetime increases first with increasing $F$ 
until a certain threshold, and then decreases with increasing $F$ similar to a slip bond. The catch-bond behavior was first predicted
theoretically \cite{Dembo_RLK_1988} and later discovered for leukocytes experimentally \cite{Marshall_DOC_2003}. The parallel bond-cluster 
model for slip bonds has also been adapted to the case of catch bonds \cite{Sun_ELC_2012,Sun_ELC_2012,Novikova_CFB_2013}. 

\begin{figure}[t!]
	\centering 
	\includegraphics[width=0.8\columnwidth]{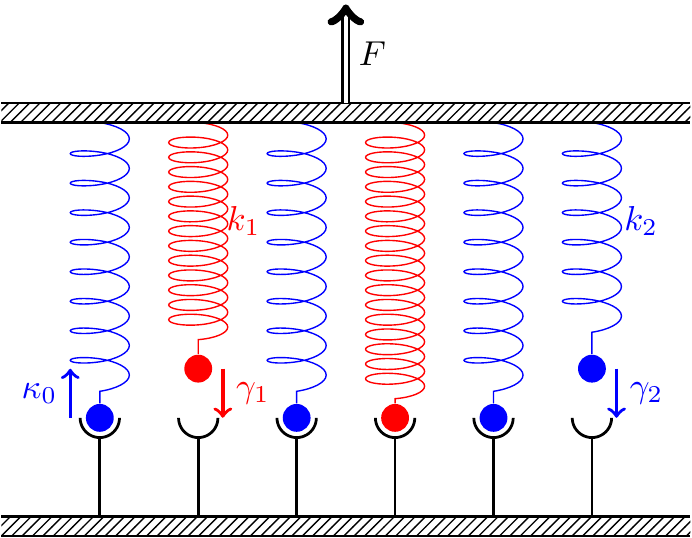}
	\caption{Heterogeneous parallel bond cluster with two different bond populations indicated by blue and red colors.
	$\gamma_1$ and $\gamma_2$ are rebinding rates of the 1st and 2nd bond types, respectively. Similarly, $k_1$ and $k_2$ 
	are the corresponding spring rigidities. $\kzero$ is a reference off-rate whose inverse $1/ \kzero$ sets a basic timescale.}
	\label{fig1}
\end{figure}

In this article, we extend the homogeneous parallel-bond-cluster model to multiple bond populations and show that the stability and lifetime 
of a heterogeneous cluster can be tuned by changing the fractions of different bond populations and their extensional rigidity and binding properties. 
We use both stochastic simulations and the mean-field approach to construct critical rupture-force diagrams for a number of relevant parameters. 
Under a gradual increase of the applied load, the dissociation of a heterogeneous bond cluster can well be described by a multistep discrete process, 
which starts with the rupture of one of the bond sub-populations and continues with similar rupture steps of other sub-populations 
or shows a sudden detachment of the remaining cluster. This cluster-dissociation behavior is qualitatively independent of 
involved bond types, including slip and catch bond sub-clusters. To maximize the critical rupture force of a heterogeneous cluster
for fixed bond kinetics and sub-cluster fractions, the distribution of loads on distinct bond sub-clusters has to be such that individual 
loads on the sub-clusters are equal to their critical rupture forces. The load balance within a heterogeneous cluster is controlled by the ratio of bond 
extensional rigidities. Our results show that a strong load disbalance (i.e. majority of the total force is applied on a single sub-cluster) 
is generally disadvantageous for overall cluster stability, because the bond sub-cluster carrying the load majority quickly ruptures 
without significant stability enhancement. Finally, we employ master equation to compute the lifetime of heterogeneous clusters for 
various parameters, and show that the heterogeneity can drastically affect cluster lifetime.

\section{Methods and models}

\subsection{Parallel bond cluster model}

We start with the original model for parallel slip-bond cluster 
under a constant loading \cite{Bell_MSA_1978,Erdmann_SAC_2004}. The system contains $\Nt$ adhesion sites, where $N(\tau)\le\Nt$
($\tau$ is time) bonds or bound springs with an extensional rigidity $k$ and a dimensionless rebinding rate $\gamma =\kappa^{\rm on}/ \kzero$ 
can stochastically form under an external force $F$. From the stability analysis \cite{Bell_MSA_1978}, with the assumption that each 
spring shares the same force $F/N$, there exists a critical force $\fc$ below which the cluster equilibrates to an average number of 
bonds $\langle N \rangle$ and above which the cluster is unstable and dissociates, i.e. $N=0$. The critical force $\fc$ and 
the critical number $\Nc$ of springs for slip-bond cluster are given by \cite{Bell_MSA_1978,Erdmann_SDA_2004}
\begin{equation}
\label{fc}
\frac{\fc}{\fd} = \Nt\,\text{pln}(\gamma/e), \,\,\,\,\,\,\,\, \Nc = \Nt\dfrac{\text{pln}(\gamma/e)}{1+\text{pln}(\gamma/e)},
\end{equation}
where $\text{pln}(a)$ is the product logarithm function which solves the equation $x\,e^x=a$.

We generalize this model to a heterogeneous cluster with two different bond populations, characterized by the extensional rigidities $k_1$ and $k_2$ 
and the rebinding rates $\gamma_1 = \kappa^{\rm on}_1/ \kzero$ and $\gamma_2 = \kappa^{\rm on}_2/ \kzero$, see Fig.~\ref{fig1}. 
Despite the fact that unstressed off-rates of distinct slip-bond populations can be different, the rebinding rates $\gamma_1$ and $\gamma_2$ 
are defined here using a reference $\kzero$ off-rate, whose inverse $1/ \kzero$ sets the basic timescale in the system. The total number $\Nt$ of adhesion sites is assumed to be 
constant, and $\rho$ determines a fraction of type-1 adhesion sites, such that $\Nt_1 = \rho \Nt$ and $\Nt_2 = (1 -\rho) \Nt$.
We also define a spring-rigidity ratio $\kr=k_1/k_2$ and a rebinding ratio $\gratio=\gamma_1/\gamma_2$. 
At any time $\tau=t \kzero$, the applied force is
\begin{equation}
F = N_1 k_1\Delta x + N_2 k_2\Delta x = \left(N_1\kr+N_2\right)k_2\Delta x,
\end{equation}
where $\Delta x$ is the extension of bound springs. $f_1$ and $f_2$ are the forces acting on the corresponding populations 
of bond types 1 and 2, given by  
\begin{equation}
\label{f1}
f_1 = \dfrac{N_1 F \kr}{N_1\kr+N_2}, \,\,\,\,\,\, f_2 = \dfrac{N_2 F}{N_1\kr+N_2},
\end{equation}
such that $f_1+f_2=F$. Clearly, $\kr$ directly controls the distribution of forces between the two bond populations. 
For the case of $\gratio=1$ and $\kr=1$, the heterogeneous cluster becomes identical to the homogeneous bond cluster considered 
previously. 

\subsection{Mean-field approximation}

The average number of bonds is governed by two non-dimensionalized rate equations:
\begin{eqnarray}
\label{eqdeter}
\dfrac{\der N_1}{\der\tau} &= -N_1\dfrac{\kappa^{\rm off}_1}{\kzero} + (\Nt_1-N_1)\gamma_1, \\
\label{eqdeter1}
\dfrac{\der N_2}{\der\tau} &= -N_2\dfrac{\kappa^{\rm off}_2}{\kzero} + (\Nt_2-N_2)\gamma_2. 
\end{eqnarray}
For a cluster with two slip-bond populations, $\kappa^{\rm off}_1(f_1/N_1) = \kzero_1 \exp{[f_1/(N_1 \fd_1)]}$ and 
$\kappa^{\rm off}_2(f_2/N_2)  = \kzero_2 \exp{[f_2/(N_2 \fd_2)]}$.
$\fd_1$ and $\fd_2$ are the two force scales which are assumed to be the same, $\fd_1=\fd_2=\fd$. 
Equations~(\ref{eqdeter}) and (\ref{eqdeter1}) are coupled via the forces $f_1$ 
and $f_2$ in Eq.~(\ref{f1}) and are used to deduce the average number $\meanN = \langle N_1 + N_2\rangle$ of bonds and 
critical force $\fc$ of the entire cluster. Note that further on all quantities with force dimensions are implicitly normalized by $\fd$.  

\subsection{Master equation}
An extension of the mean-field approach is a one-step, two-variable master equation 
for this system \cite{Novikova_ERD_2019}. If $P_{i,j}$ is the probability of $i$ type-1 bonds and $j$ 
type-2 bonds, then the master equation is
\begin{multline}
\label{ME}
\dfrac{\der P_{i,j}}{\der\tau} = r_1^{i+1,j}P_{i+1,j}+r_2^{i,j+1}P_{i,j+1} +g_1^{i-1,j}P_{i-1,j}+ \\ 
 g_2^{i,j-1}P_{i,j-1} -\left[r_1^{i,j}+r_2^{i,j}+g_1^{i,j}+g_2^{i,j}\right]P_{i,j},   
\end{multline}
where $r_1^{i,j}=i\, \kappa^{\rm off}_1(f_1/i) / \kzero$ and $r_2^{i,j}=j\, \kappa^{\rm off}_2(f_2/j) / \kzero$ 
are the reverse rates and $g_1^{i,j}=\gamma_1(\Nt_1 - i)$ and $g_2^{i,j}=\gamma_2(\Nt_2 - j)$ are the rebinding 
rates for type-1 and type-2 bonds, respectively. Then, the average numbers of type-1 and type-2  bonds can be computed as 
\begin{equation}
\label{avg}
\avg{i} = \sum_{i,j}iP_{i,j}, \,\,\,\,\,\, \avg{j} = \sum_{i,j}jP_{i,j}. 
\end{equation} 

Cluster lifetime $T_{i,j}$ corresponds to the time required for the cluster to dissociate completely, i.e. time for 
$\{i,j\}\rightarrow \{0,0\}$ (no bonds). The equation for $T_{i,j}$ of a heterogeneous cluster can be derived similarly 
to the approach for parallel cluster with a single bond type \cite{Van_Kampen_SPP_1992}, and then solved numerically, 
following the recursive scheme for the lifetime of a cluster with catch and slip bonds \cite{Novikova_ERD_2019}. 
For $i$ type-1 bonds and $j$ type-2 bonds, after a time interval $\Delta t$, 
the cluster jumps to one of the four neighboring states ($\{i+1,j\}$, $\{i-1,j\}$, $\{i,j+1\}$, or $\{i,j-1\}$) 
or it remains at the state $\{i,j\}$, which is described as 
\begin{align}
T_{i,j} -\Delta t  &= g_1^{i,j} \Delta t\,T_{i+1,j}+g_2^{i,j} \Delta t\,T_{i,j+1}+ r_1^{i,j} \Delta t\,T_{i-1,j} \nonumber \\ 
& +r_2^{i,j} \Delta t\,T_{i,j-1}  \\ 
&+\left( 1-g_1^{i,j}\Delta t-g_2^{i,j}\Delta t- r_1^{i,j}\Delta t-r_2^{i,j}\Delta t\right) T_{i,j}.  \nonumber
\end{align}
This equation can be simplified to
\begin{align}
\label{finalT}
-1 &= g_1^{i,j}\left(T_{i+1,j}-T_{i,j}\right) + g_2^{i,j}\left(T_{i,j+1}-T_{i,j}\right) \nonumber\\ &+ 
r_1^{i,j}\left(T_{i-1,j}-T_{i,j}\right) + r_2^{i,j}\left(T_{i,j-1}-T_{i,j}\right).
\end{align}
This recursive relation generates $\Nt_1\times \Nt_2$ algebraic equations, which are solved by inverting the corresponding 
coefficient matrix \cite{Van_Kampen_SPP_1992,Novikova_ERD_2019}. The computation also requires appropriate boundary conditions 
for on- and off-rates for both bond types and the state $\{0,0\}$ acts as an absorbing boundary condition, i.e. $T_{0,0}=0$. 

\begin{figure}[t!]
	\centering
	\includegraphics[width=1.0\columnwidth]{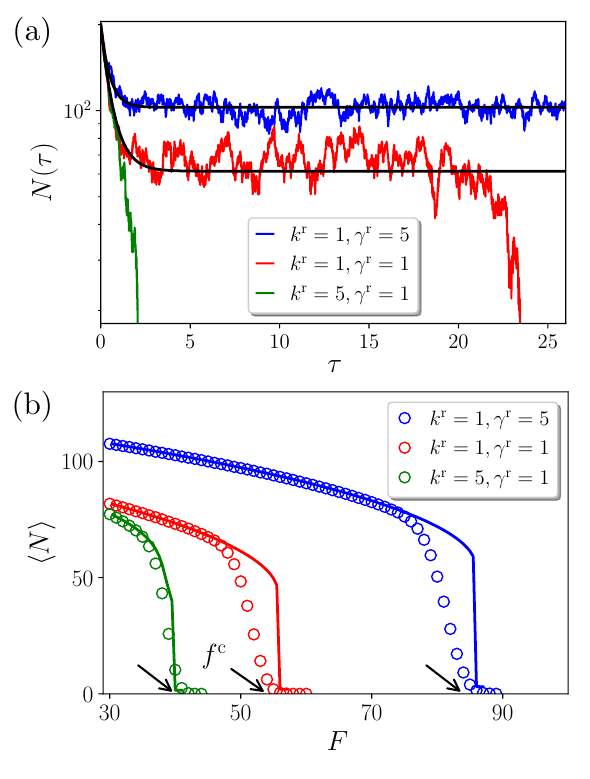}
	\caption{Behavior of a heterogeneous cluster with two slip-bond populations. (a) Evolution of $N(\tau) = N_1(\tau) + N_2(\tau)$ for $F=50$, $\rho =0.3$, and $\Nt=200$. Trajectories from stochastic
	simulations are shown by colored lines, the numerical solution of Eqs.~(\ref{eqdeter}) and (\ref{eqdeter1}) by black lines. 
	The critical force for a uniform cluster with $\kr=1$ and $\gratio=1$ is $\fc\simeq 55.7$. (b) Average number $\meanN$ of bonds 
	as a function of the applied force $F$ for different $\gratio$ and $\kr$. Lines correspond to the mean-field approximation, 
	symbols to stochastic simulations. The arrows indicate the corresponding $\fc$ values.}
	\label{fig2}
\end{figure}  

\subsection{Stochastic simulations}
Another approach for analyzing cluster stability is direct stochastic simulations using 
the Gillespie's algorithm~\cite{Gillespie_ESS_1977}. The heterogeneous system is described by four rate equations, two for each type 
of bonds, representing their association and dissociation as
\begin{align}
\label{rea1}
O_1 & \xrightarrow[]{\kappa^{\rm on}_1}C_1,	\\ 
\label{rea2}
C_1 & \xrightarrow[]{\kappa^{\rm off}_1}O_1, 	\\
\label{rea3}
O_2 & \xrightarrow[]{\kappa^{\rm on}_2}C_2, 	\\
\label{rea4}
C_2 & \xrightarrow[]{\kappa^{\rm off}_2}O_2,  
\end{align} 
where $O$ denotes an open state and $C$ a closed state. Here, $\kappa^{\rm on}_1$, $\kappa^{\rm off}_1$, $\kappa^{\rm on}_2$, and 
$\kappa^{\rm off}_2$ are the reaction rates.    

The algorithm follows by generating two independent random numbers $\xi_1$ and $\xi_2$ that are uniformly distributed at the interval $[0,1]$. 
Then, $\xi_1$ is employed to define the time $\mathrm{d}\tau$ at which next reaction occurs, while $\xi_2$ is used to choose which reaction occurs next. 
The time $\mathrm{d}\tau$ is given by
\begin{equation}
\mathrm{d}\tau = \dfrac{1}{\alpha}\log\left(\dfrac{1}{\xi_1}\right),
\end{equation}
where $\alpha = \alpha_1+\alpha_2+\alpha_3+\alpha_4$ is the combined propensity function with separate propensity functions
\begin{align}
\alpha_1 &= \gamma_1\,(\Nt_1 - N_1),	\\
\alpha_2 &= N_1\,\kappa^{\rm off}_1(f_1/N_1) / \kzero,	\\
\alpha_3 &= \gamma_2\,(\Nt_2 - N_2),	\\
\alpha_4 &= N_2\,\kappa^{\rm off}_2(f_2/N_2) / \kzero.	
\end{align}
For $\xi_2 \in [0,\alpha_1/\alpha)$ the first reaction [Eq.~(\ref{rea1})], for $\xi_2 \in [\alpha_1/\alpha,
(\alpha_1+\alpha_2)/\alpha)$ the second reaction [Eq.~(\ref{rea2})], for $\xi_2 \in [(\alpha_1+\alpha_2)/\alpha, 
(\alpha_1+\alpha_2+\alpha_3)/\alpha)$ the third reaction [Eq.~(\ref{rea3})], and for $\xi_2 \in [(\alpha_1+\alpha_2+\alpha_3)/\alpha,1]$
the fourth reaction [Eq.~(\ref{rea4})] is chosen, respectively. Following this algorithm, $N_1(\tau)$ and $N_2(\tau)$ are 
advanced in time until the entire cluster detaches (i.e. $N_1=0$ and $N_2=0$) or when a pre-defined maximum number of simulation steps 
is reached.
 
\section{Results}

We first consider a heterogeneous cluster with two slip-bond populations denoted as slip-slip bond cluster.  
For simplicity, we assume $\kzero_1 = \kzero_2 = \kzero$.
Figure \ref{fig2}(a) shows typical evolution of $N(\tau)= N_1(\tau) + N_2(\tau)$ for several slip-slip bond clusters with various $\kr$ and $\gratio$,
where $F=50$, $\rho=0.3$, $\Nt=200$. The case of $\kr=1$ and $\gratio=1$ corresponds to a homogeneous cluster for which 
$\fc\simeq 55.7$. Even though $F<\fc$, the stochastic trajectory (red line) shows a complete cluster dissociation due to fluctuations 
in $N$ and the condition of $N(\tau)=0$ for simulation termination. Note that cluster dissociation occurs more frequently when 
the applied force is approaching $\fc$. The corresponding solution of deterministic Eqs.~(\ref{eqdeter}) and (\ref{eqdeter1}) shown 
by the black line converges to a constant $N$ for large $\tau$. For $\kr=1$ and $\gratio =5$ (blue line), the cluster is very 
stable because the critical force is much larger than $F=50$, which is evident from Fig.~\ref{fig2}(b), where 
the average number $\meanN$ of bonds is presented as a function of $F$ for different $\gratio$ and $\kr$. In contrast, the cluster 
with $\kr=5$ and $\gratio=1$ quickly dissociates at $F=50$, as it significantly exceeds the critical force. The differences 
in $\meanN$ between stochastic simulations (symbols) and deterministic solutions (lines), as $F$ approaches $\fc$ in Fig.~\ref{fig2}(b), 
characterize the fraction of simulations where cluster dissociation has occurred within the total simulation time. Note that the cluster 
lifetime is always finite, but it can be much larger than the total time of stochastic simulations when the applied force is considerably 
smaller than $\fc$.        

\subsection{Critical force and stability enhancement}

Dissociation of a heterogeneous cluster can be thought of as a multistep process. For two bond populations, as the applied 
force $F$ is increased, one of the sub-clusters dissociates first, followed by the detachment of the other. Thus, depending 
on how $F$ is shared between two sub-clusters which is controlled by $\kr$ [see Eq.~(\ref{f1})], there exist two possibilities
\begin{eqnarray}
\label{fc1}
\textrm{(i)} \,\,\, f_1 & = & \fc_1, \,\, N_1=\Nc_1 \,\, \& \,\, f_2 \leq \fc_2, \\
\label{fc2}
\textrm{(ii)} \,\,\, f_2 & = & \fc_2, \,\, N_2=\Nc_2 \,\, \& \,\, f_1 \leq \fc_1,
\end{eqnarray}
where $\fc_1$, $\fc_2$, $\Nc_1$, and $\Nc_2$ are the corresponding 
critical forces and numbers of bonds of the two sub-clusters separately.
In the first case, Eqs.~(\ref{fc1}) and (\ref{eqdeter1}) with two unknowns $F$ and $N_2$ become
\begin{eqnarray}
\dfrac{\Nc_1 F \kr}{\Nc_1\kr+N_2} & = & \fc_1, \\
((1-\rho)\Nt - N_2)\gamma_2 & = & N_2\, \kappa^{\rm off}_2 (f_2/N_2) / \kzero, 
\end{eqnarray}
from which the applied force $F^{\rm c}$ required to initially dissociate the first sub-cluster can be computed.
In the second case, Eqs.~(\ref{fc2}) and (\ref{eqdeter}) with two unknowns $F$ and $N_1$ are given by
\begin{eqnarray}
\dfrac{\Nc_2 F}{N_1\kr+\Nc_2} & = & \fc_2, \\
(\rho\Nt - N_1)\gamma_1 & = & N_1 \, \kappa^{\rm off}_1(f_1/N_1) / \kzero, 
\end{eqnarray}
from which the applied force $F^{\rm c}$ required to initially dissociate the second sub-cluster can be calculated
numerically. 
 
After one of the sub-clusters has initially dissociated under the force $F^{\rm c}$, the other sub-cluster can immediately 
rupture, if  $F^{\rm c}$ is larger than or equal to its individual critical force, or detachment of the remaining sub-cluster 
requires a force that is larger than $F^{\rm c}$. This condition can be taken into account through the requirement that 
the critical force $\fc$ for rupturing the entire cluster must necessarily satisfy $\fc \geq \fm = \textrm{max}(\fc_1,\fc_2)$. 
Thus, we obtain $\fc = F^{\rm c}$ if $F^{\rm c} \geq \fm$, and $\fc = \fm$ otherwise.   

\begin{figure}[t!]
	\centering
	\includegraphics[width=1.0\columnwidth]{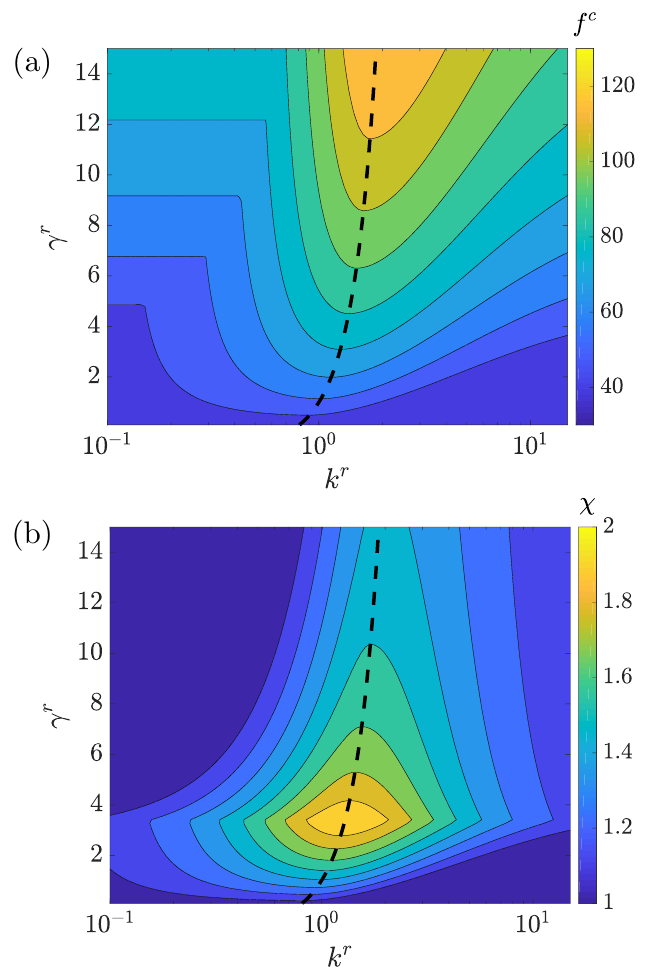}
	\caption{Stability characteristics of a slip-slip bond cluster. (a) $\fc$ map for different $\kr$ and $\gratio$. (b) Cluster stability enhancement 
     $\chi=\fc/\fm$ by a weaker sub-cluster for various $\kr$ and $\gratio$. The dashed lines show $\kr_\textrm{opt}$ 
		values from Eq.~(\ref{optkr}), which correspond to a maximum $\fc$ for a fixed $\gratio$. Here, $\rho=0.3$ and $\Nt=200$.}
	\label{fig3}
\end{figure}

Figure~\ref{fig3}(a) shows the critical force map as a function of $\kr$ and $\gratio$ for a slip-slip bond cluster with $\rho=0.3$.
Note that for any fixed $\gratio$, there exists a maximum $\fc$, which corresponds to the special case with $f_1=\fc_1$ \& 
$f_2=\fc_2$. Then, the ratio  
\begin{equation}
\dfrac{f_1}{f_2} = \dfrac{\fc_1}{\fc_2} = \dfrac{\Nc_1 \kr}{\Nc_2} 
\end{equation}
allows the calculation of optimal $\kr$ values for cluster stability. For a slip-slip bond cluster, we obtain   
\begin{equation}
\label{optkr}
\kr_\textrm{opt} = \frac{1+\textrm{pln}(\gamma^\prime_1/e)}{1+\textrm{pln}(\gamma^\prime_2/e)},
\end{equation}    
where $\gamma^\prime_1 = \gamma_1 \kzero / \kzero_1$ and $\gamma^\prime_2 = \gamma_2 \kzero / \kzero_2$. 
Thus, for a fixed $\gratio$, the largest $\fc$ is achieved when the forces on individual bond sub-clusters, which is controlled by $\kr$, 
are equal to the corresponding critical forces $\fc_1$ and $\fc_2$. Surprisingly, $\kr_\textrm{opt}$ for a slip-slip bond cluster 
depends only on the rebinding rates, and is independent of $\rho$. The dashed line in Fig.~\ref{fig3}(a) represents $\kr_\textrm{opt}$ 
and separates the $\fc$ map into two regions. Region on the right side from the dashed line corresponds to Eq.~(\ref{fc1}), 
where the first sub-cluster dissociates first. Consequently, the region on the left side corresponds to Eq.~(\ref{fc2}), where the second 
sub-cluster is initially ruptured. Noteworthy, $\kr_\textrm{opt}$ is a weakly increasing function of $\gratio$, indicating that large or small 
values of $\kr$ (or strongly disproportionate load sharing between sub-clusters) are disadvantageous for the stability of entire cluster. 

It is also interesting to consider the quantity 
\begin{equation}
\chi = \dfrac{\fc}{\textrm{max}(\fc_1,\fc_2)} = \dfrac{\fc}{\fm}, 
\end{equation}
which describes the effect of a weaker sub-cluster (i.e. the sub-cluster with a smaller critical force $\textrm{min}(\fc_1,\fc_2)$) 
on $\fc$ in comparison with the critical force $\textrm{max}(\fc_1,\fc_2)$ of the strongest sub-cluster. Figure \ref{fig3}(b) shows 
the stability enhancement $\chi$ by a weaker sub-cluster as a function of $\kr$ and $\gratio$. It can be shown analytically (see Appendix~\ref{sec:app}) 
that the maximum possible enhancement is $\chi_\mathrm{max} = 2$, which is located on the $\kr_\textrm{opt}$ line at a $\gratio$ value
determined by the equality $\fc_1=\fc_2$. Thus, maximum enhancement of $\fc$ by a weaker sub-cluster is achieved when critical 
forces of individual sub-clusters are equal. Furthermore, large or small values of $\gratio$ and $\kr$ (compared to unity) generally result in $\chi \approx 1$,
and therefore, nearly no stability enhancement by the addition of a weaker sub-cluster. 

\begin{figure}[hbt!]
	\centering
	\includegraphics[width=1.0\columnwidth]{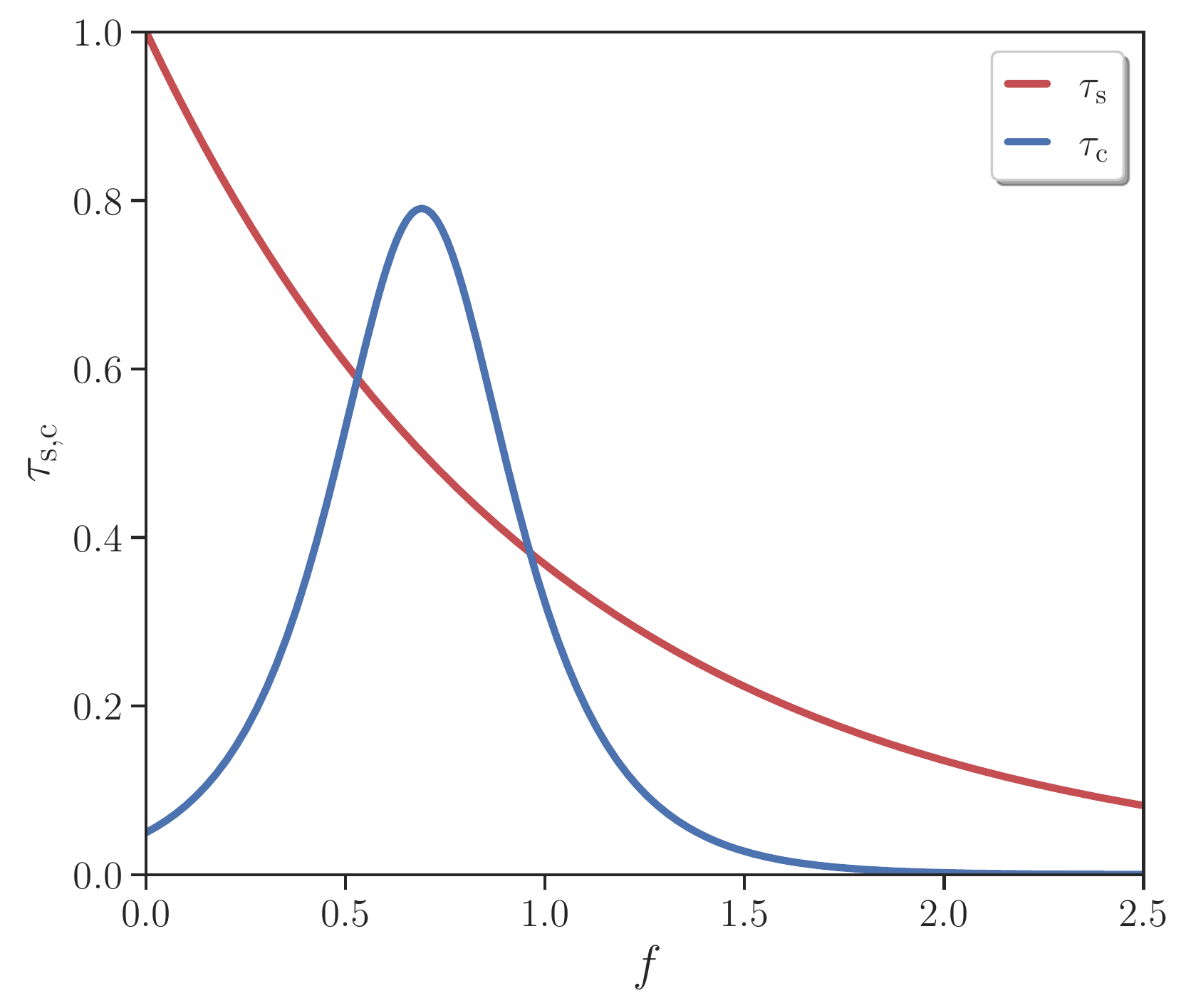}
	\caption{Comparison of  lifetimes of single catch ($\tau_{\rm c}$) and slip ($\tau_{\rm s}$) bonds defined as $\kzero/\kappa^{\rm off}$. 
		Here, $\kzero_{\rm s} / \kzero = 0.02$,  $\kzero_{\rm c} / \kzero = 20$, $\beta = 0.2$, and $\kzero_2 / \kzero = 1$.}
	\label{fig4}
\end{figure}

\subsection{Catch-slip bond cluster}

To study the effect of a catch-bond sub-population on cluster stability, we consider a catch-slip bond system where the type-1 bond 
population is represented by catch bonds and type-2 by slip bonds. The off-rate of catch bonds is given by \cite{Pereverzev_TPM_2005}
\begin{equation}
	\label{catch_or}
	\kappa^{\rm off}_1(f_1/N_1) = \kzero_{\rm s} e^{f_1/(N_1 \fd \beta)} + \kzero_{\rm c} e^{ - f_1/(N_1 \fd \beta)},  
\end{equation}
where $\kzero_{\rm s}$ and $\kzero_{\rm c}$ are unstressed off-rates of the slip and catch contributions to $\kappa^{\rm off}_1$, 
and $\beta$ is a non-dimensional quantity that alters the characteristic force scale (i.e. $\fd_1 = \fd \beta$). To illustrate differences between catch 
and slip bonds, Fig.~\ref{fig4} shows the lifetimes of single bonds defined as $\kzero/\kappa^{\rm off}$. Lifetime of the slip bond 
$\tau_{\rm s}$ is a monotonically decreasing function of force $f$, while for the catch bond, $\tau_{\rm c}$ first increases and then 
decreases, representing catch and slip parts.  

\begin{figure}[hbt!]
	\centering
	\includegraphics[width=1.0\columnwidth]{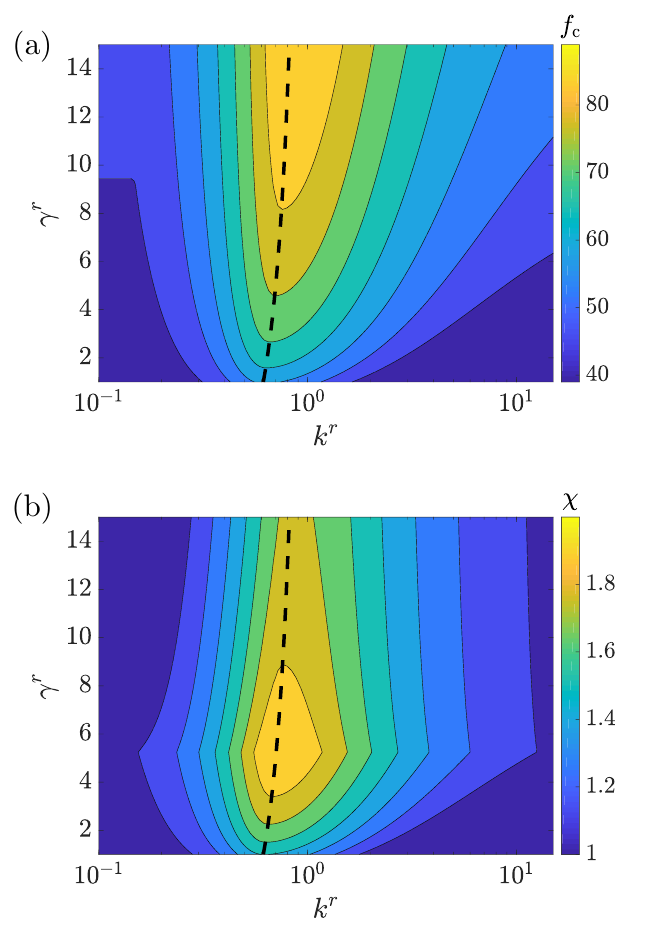}
	\caption{Cluster stability for a catch-slip bond cluster. (a) $\fc$ map and (b) cluster stability enhancement 
		$\chi$ by a weaker sub-cluster for different $\kr$ and $\gratio$ values. The dashed lines correspond to 
		$\kr_\textrm{opt}$ calculated numerically. Here, $\rho=0.3$, $\Nt=200$, and other parameters are the same as 
		in Fig.~\ref{fig4}.}
	\label{fig5}
\end{figure}

The critical force of a catch-slip bond cluster is computed similarly to that of the slip-slip bond cluster by using Eq.~(\ref{fc1}) or 
Eq.~(\ref{fc2}), depending on whether $\kr$ is larger or smaller than $\kr_\textrm{opt}$. This computation of $\fc$ first requires 
the calculation of $\fc_1$ and $\Nc_1$ of the catch-bond sub-cluster, which is performed by solving  a system of the rate 
equation and its derivative, i.e. $\dot{N_1}=0$ and $\mathrm{d}\dot{N_1}/\mathrm{d}N_1=0$. Thus, for each pair of $\kr$ 
and $\gratio$, $\fc_1$ and $\Nc_1$ are first pre-computed in order to find the optimal $\kr$ value as 
$\kr_\textrm{opt}=(\fc_1/\fc_2)/(\Nc_1/\Nc_2)$, which is then used to determine whether the condition in Eq.~(\ref{fc1}) or 
Eq.~(\ref{fc2}) has to be considered. Figure~\ref{fig5} presents critical force $\fc$ and stability enhancement $\chi$ maps of a catch-slip bond 
cluster as a function of $\kr$ and $\gratio$. Both $\fc$ and $\chi$ maps resemble the corresponding contour plots for the slip-slip bond 
cluster in Fig.~\ref{fig3}. Therefore, the multistep rupture behavior discussed for slip-slip clusters above is also directly applicable for 
catch-slip bond clusters. 

\begin{figure*}[htb!]
	\centering
	\includegraphics[width=1.0\textwidth]{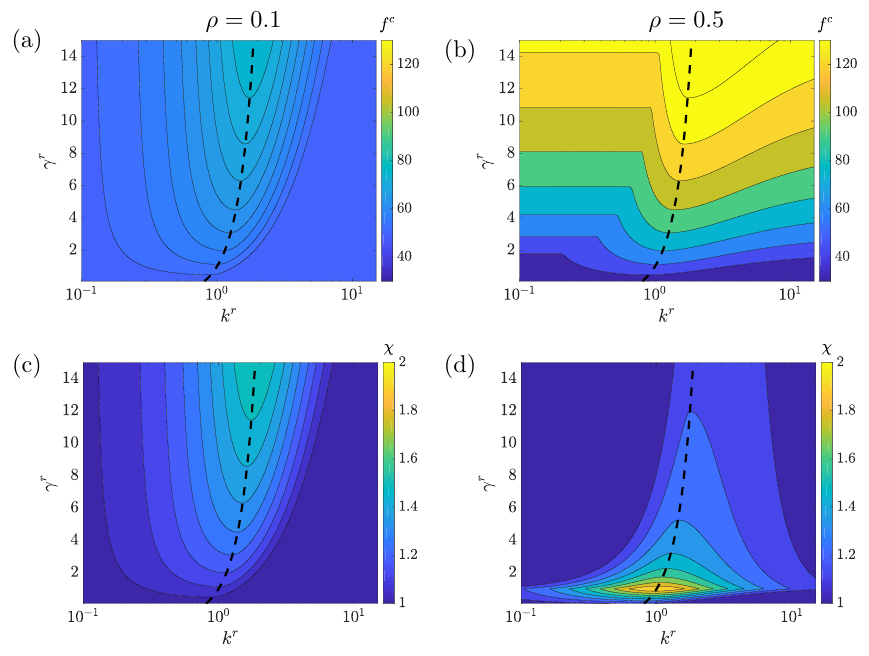}
	\caption{[(a) \& (b)] Critical force $\fc$ and [(c) \& (d)] stability enhancement $\chi$ of a slip-slip bond cluster for different bond fractions [(a) \& (c)] $\rho=0.1$ and [(b) \& (d)] $\rho=0.5$
		with $\Nt=200$. Note that the ranges of $\kr$ and $\gratio$ for $\rho=0.1$ are different from those for $\rho=0.5$ and $\rho=0.3$ in Fig.~\ref{fig3}. 
		The dashed lines indicate $\kr_\textrm{opt}$ values from Eq.~(\ref{optkr}), which correspond to a maximum $\fc$ for a fixed $\gratio$.}
	\label{fig6}
\end{figure*}

\subsection{Effect of bond-population fraction}

Figure \ref{fig6} shows the critical force $\fc$ and stability enhancement $\chi$ of a slip-slip bond cluster for two fractions ($\rho=0.1$ and
$\rho=0.5$) of the first bond population. For $\rho=0.1$, $\fc$ values are smaller than those for $\rho=0.3$ in Fig.~\ref{fig3}, while for $\rho=0.5$, 
$\fc$ values are larger than for $\rho=0.3$. However, the overall structure of the $\fc$ and $\chi$ maps remains qualitatively similar for all $\rho$ 
values. Note that the value of $\chi_\mathrm{max} = 2$ is also independent of $\rho$, but its position on the $\kr_\textrm{opt}$ line changes 
with $\rho$.  For example, $\chi_\mathrm{max}$ lies at $\gratio\simeq82$  for $\rho=0.1$, at $\gratio\simeq 3.5$ for $\rho=0.3$, and at exactly 
$\gratio=1$ for $\rho=0.5$, which corresponds to a homogeneous cluster.

\begin{figure}[htb!]
	\centering
	\includegraphics[width=1.0\columnwidth]{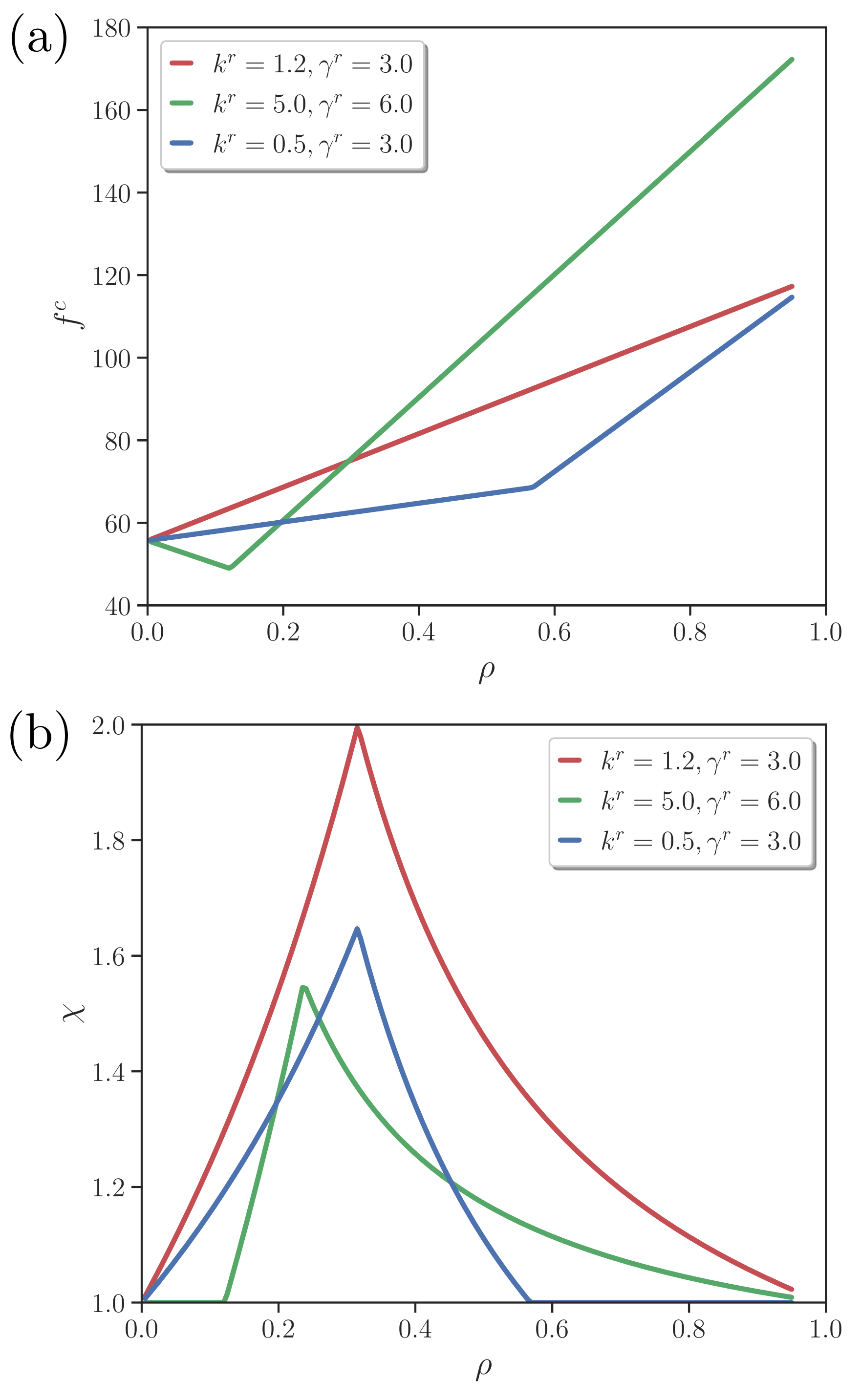}
	\caption{Effect of bond fraction $\rho$ on (a) the critical force $\fc$ and (b) the stability enhancement $\chi$ of a slip-slip bond cluster for several 
		$\gratio$ and $\kr$ values. Here, $\Nt=200$.}
	\label{fig7}
\end{figure}  

For a fixed $\gratio$, the optimal fraction $\rho_\mathrm{opt}$, such that $\chi_\mathrm{max}$ lies at $\gratio$, can be found from the equality 
$\fc_1=\fc_2$ that for a slip-slip bond cluster yields (see Appendix~\ref{sec:app})   
\begin{equation}
\rho_\mathrm{opt} = \frac{\textrm{pln}(\gamma^\prime_2/e)}{\textrm{pln}(\gamma^\prime_1/e)+\textrm{pln}(\gamma^\prime_2/e)}.
\label{rho_opt}
\end{equation}     
Thus, the fraction $\rho$ can also be tuned to control $\fc$ and $\chi$, which is illustrated in Fig.~\ref{fig7} for different combinations of $\kr$ and 
$\gratio$. For the same $\gratio$, two curves with different $\kr$ values (red and blue curves) give rise to maximum stability enhancement at the same 
fraction because $\rho_\mathrm{opt}$ is independent of $\kr$, which is consistent with no dependence of $\kr_\textrm{opt}$ on $\rho$ in Eq.~(\ref{optkr}).  
However, only $\kr$ values close to $\kr_\textrm{opt}$ approach $\chi_\mathrm{max} = 2$. In case of three different bond populations, 
$\chi_\mathrm{max} = 3$ (see Appendix~\ref{sec:app}), suggesting that $\chi_\mathrm{max}$ is equal to the number of bond populations.
   
\begin{figure}[t!]
	\centering
	\includegraphics[width=1.0\columnwidth]{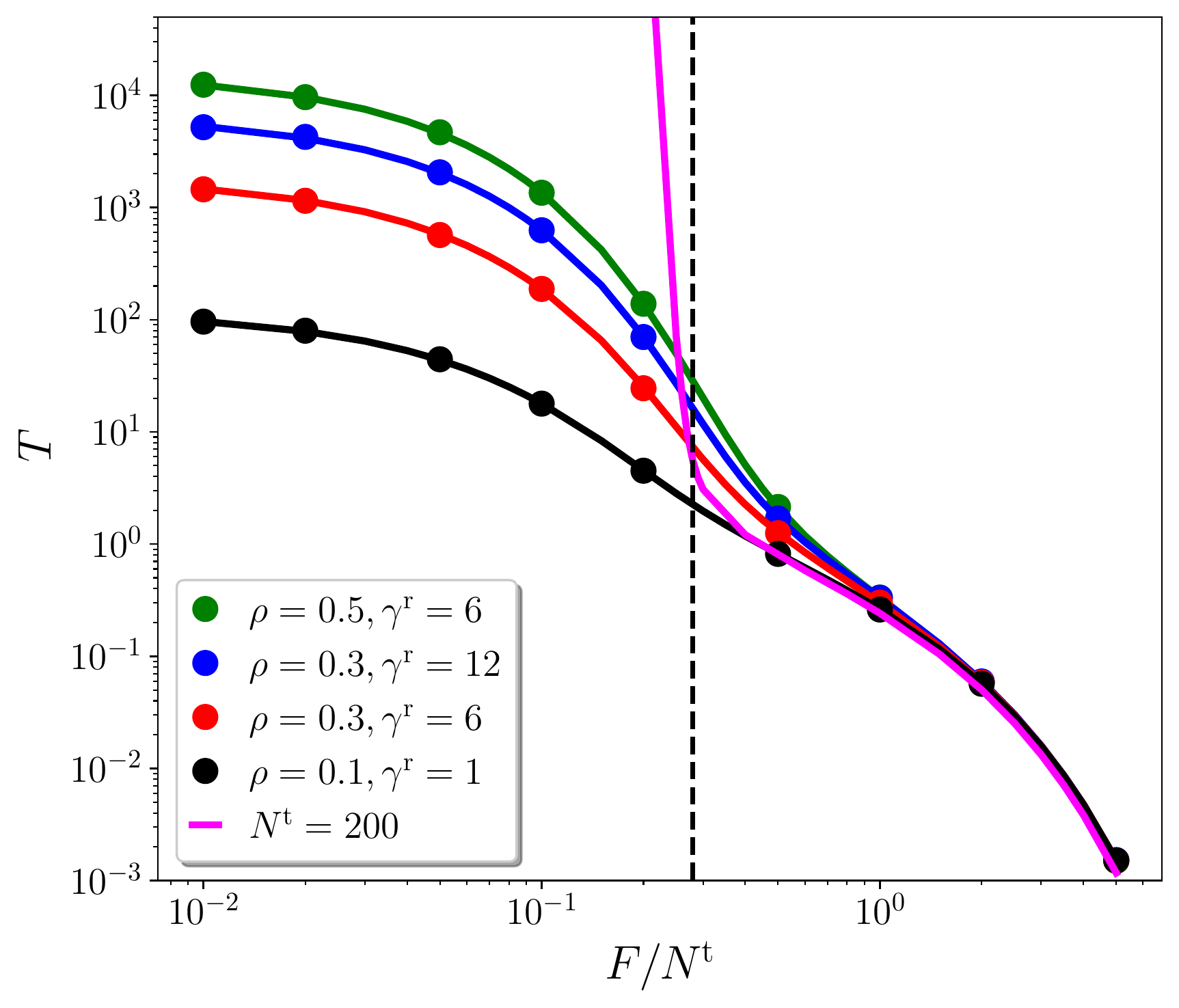}
	\caption{Lifetime of a slip-slip bond cluster for different $\gratio$ and $\rho$ with $\kr=1$ and $\Nt=10$. Circle symbols represent 
	stochastic simulations and solid lines are obtained using the master equation (\ref{ME}). The dashed line marks the critical 
	force for $\rho=0.1$ and $\gratio=1$. The magenta line is for $\Nt=200$ and $\gratio=1$. Above the critical force, $T\sim \exp{(-F/\Nt)}$.}
	\label{fig8}
\end{figure} 

\subsection{Cluster lifetime}

For a single bond with vanishing rebinding rate, its lifetime $T$ is simply  
$T = 1/\kappa_{\rm off}$. In the mean-field description, when the applied force is less than $\fc$, 
the cluster lifetime is infinite. However, stochastic fluctuations may result in cluster dissociation within a finite time. 
The lifetime of a heterogeneous cluster can be obtained from the master equation (\ref{ME}), and also directly from stochastic 
simulations. Figure~\ref{fig8} presents lifetimes of slip-slip bond clusters for different model parameters, with an excellent agreement between 
the results from Eq.~(\ref{ME}) and stochastic simulations for $\Nt=10$. Clearly, cluster lifetimes are finite even when $F<\fc$. 
$T$ increases drastically with increasing $\gratio$ for a given $\rho$. Furthermore, the cluster lifetime strongly increases 
with increasing $\rho$ for a fixed $\gratio > 1$. Differences in $T$ for various $\kr$ can be pronounced when the applied force $F$
is close to $\fc$, since a change in the load distribution controlled by $\kr$  can make an initially stable cluster unstable and vice versa. 
However, for $F \ll \fc$ and  $F \gg \fc$, the effect of $\kr$ on $T$ is expected to be negligible, as load re-distribution will 
have no significant effect on whole cluster stability or rupture. Note that we have employed a relatively small value of $\Nt=10$ because the lifetime increases 
exponentially with $\Nt$, so that direct stochastic simulations do not permit the calculation of $T$ for large $\Nt$. 
Figure \ref{fig8} also shows that the lifetime for $\Nt=200$ (magenta color) rapidly increases when the applied force becomes 
smaller than the critical force.

\section{Discussion and conclusions} 

In this study, we have investigated the stability of heterogeneous bond clusters under a constant load using the theoretical model
for parallel bond cluster. One of our main results is that the rupture of heterogeneous bond clusters usually occurs as a multistep 
discrete process. As the load on a cluster with two bond populations is gradually increased, one of the sub-clusters eventually 
ruptures when the individual load on this sub-cluster exceeds its critical rupture force. After the first bond sub-cluster has 
dissociated, the total applied load $F$ fully acts on the second remaining sub-cluster, resulting in either its immediate rupture 
if $F$ is larger than its critical rupture force or a further increase of the applied load is required to dissociate the second sub-cluster. 
This rupture behavior of a two-population cluster can straightforwardly be generalized to the dissociation of bond clusters with 
many sub-populations. The sequence of sub-cluster rupture in this process depends on the distribution of total load among 
bond sub-clusters and their individual critical rupture forces. For fixed fractions of different bond populations, the load distribution 
is directly controlled by the ratios $\kr$ of bond extensional rigidities, while the sub-cluster rupture forces are governed by 
bond kinetic rates. Furthermore, bond-population fractions within a heterogeneous cluster affect both the load distribution 
and the sub-cluster critical forces. 

The discrete multistep process of cluster dissociation described above is based on the mean-field approximation. In reality, 
stochastic fluctuations in the number of bonds of different sub-populations can affect the behavior of a heterogeneous cluster. 
For instance, the sequence, in which different bond sub-populations are ruptured as predicted by the mean-field theory, can be 
altered. Furthermore, the step-like dissociation of various bond sub-clusters in the mean-field approximation  is partially smoothed 
through the stochastic fluctuations in bond numbers, as illustrated in Fig.~\ref{fig2}(b). Note that these differences between 
the stochastic behavior of a cluster and its mean-field approximation are only expected when the applied force is sufficiently close 
to the critical force for cluster rupture. Similarly, spontaneous rupture of a bond sub-cluster can unexpectedly occur if the distributed 
load on that sub-cluster is close to its individual rupture force.  

An interesting conclusion from our study is that the multistep rupture process of a heterogeneous cluster is qualitatively independent 
whether only slip or catch bond populations are considered or their mixture. Note that in case of catch bonds, in addition to the effect of stochastic fluctuations
already discussed, such fluctuations may lead to a sub-cluster rupture at nearly zero load, since the lifetime of catch bonds without applied 
forces can be small. Therefore, catch-bond sub-populations are expected to enhance cluster stability primarily under a non-zero load.                                 

Our analysis shows that there exists an optimal spring-rigidity ratio $\kr_\textrm{opt}$ (or optimal load balancing among bond 
sub-clusters) that results in the maximum critical force for a given $\gratio$ and $\rho$. The optimal load distribution is achieved 
when the individual loads on different bond sub-clusters are equal to their corresponding critical forces, such that all sub-clusters rupture at the 
same time. This means that the 
maximum possible rupture force of the whole cluster is simply $f^{\rm max} = \sum_i^m \fc_i$, where $m$ is the number of 
different bond populations. Thus, for fixed critical forces of bond sub-clusters, we can define a range of possible critical forces 
of a heterogeneous cluster as $f^{\rm c,m} \le \fc \le f^{\rm max}$ where $f^{\rm c,m} = \max_i^m (\fc_i)$ is the maximum of critical 
forces of all bond sub-clusters. The load distribution at $\kr_\textrm{opt}$
leads to $\fc = f^{\rm max}$, while $\kr \to 0$ or $\kr \to \infty$ results in $\fc \to f^{\rm c,m}$. Therefore, a strong load disbalance 
($\kr \ll 1$ or $\kr \gg 1$) has generally no advantages for overall cluster stability, because one of the bond populations carries 
the majority of the load and gets easily ruptured without significant stability enhancement. 

The upper bound on possible critical forces of a heterogeneous cluster discussed above provides an intuitive explanation 
for the maximum stability enhancement $\chi_\mathrm{max}$ by a weaker sub-cluster. Starting from a first bond sub-cluster
with its critical rupture force $\fc_1$, we add the second bond population that is weaker than the first population, i.e. $\fc_2 \leq \fc_1$.
To maximize $f^{\rm max} = \fc_1+\fc_2$ of the whole cluster, we select $\fc_2=\fc_1$, so that $f^{\rm max} =2\fc_1 = 2\fc_2$ and 
$\chi_\mathrm{max} = 2$ for the cluster with two bond populations. In other words, the addition of a second bond population that 
is weaker than the first population can at most double the critical rupture force of the whole cluster. Clearly, this argument can be 
generalized to $m$ bond sub-populations, such that $1 \leq \chi \leq \chi_\mathrm{max}$ where $\chi_\mathrm{max} = m$.    
Note that such a maximum in $\chi$ may not easily be achieved in biological systems, as it requires simultaneous regulation 
of multiple parameters, including intrinsic properties and densities of different bond populations.    

In addition to $\gratio$ and $\kr$ for controlling $\fc$ of a heterogeneous cluster, the fraction $\rho$ of different bond populations
can be tuned to alter the critical force and lifetime of the cluster. Interestingly, $\kr_\textrm{opt}$ for a slip-slip bond cluster in 
Eq.~(\ref{optkr}) is independent of $\rho$, while the position of $\chi_\mathrm{max}$ with respect to $\gratio$ is strongly affected 
by $\rho$. Note that the ability to adjust density fractions of different bond populations has direct biological relevance as cells can 
regulate receptor density, while $\gratio$ and $\kr$ are intrinsic properties of bond populations within the cluster. 

Finally, it is important to discuss limitations of the presented theoretical model. This model is obviously too simple to quantitatively describe 
whole-cell adhesion, as it does not consider cell deformation and possible receptor and ligand mobilities. In such cases, more 
sophisticated models or simulations need to be applied, as it has been done for modeling immunological synapse 
\cite{Qi_SPF_2001,Weikl_PFA_2004} or the distribution and dynamic growth of different bond domains 
\cite{Smith_FIG_2008,Jiang_ADB_2015,Brochard_AIB_2002,Weikl_AMC_2009}.  The presented model considers heterogeneous
bond clusters in which receptor densities and population fractions are nearly conserved over time. We expect that this model is 
applicable to highly localized focal adhesion sites in order to characterize their adhesion stability. As an example, it can be applied 
to bond clusters located at the villi of leukocytes \cite{Bruehl_QLS_1996} or at the adhesive knobs of malaria-infected erythrocytes
\cite{Watermeyer_SSU_2015}. Furthermore, it is useful to estimate adhesion stability of functionalized colloidal particles used 
in functional materials \cite{Wang_DNA_2015,Zhang_DNA_2017} or for drug delivery by micro- and nano-carriers \cite{Decuzzi_ASP_2006,Cooley_IPS_2018}. 
Note that the application of this model to such systems 
requires some knowledge about involved receptor-ligand pairs, including bond kinetics and densities. Thus, the presented model 
can be used for the quantification of local cluster-adhesion measurements and better understanding of the role of different bond 
populations within heterogeneous adhesion clusters.         
  
\section*{Acknowledgments}  
We would like to thank Ulrich S. Schwarz (Heidelberg University, Germany) for insightful and stimulating discussions. 

\appendix  

\section{Optimal fraction $\rho$ and maximum stability enhancement}
\label{sec:app}

For a fixed $\gratio$, the optimal fraction $\rho_\mathrm{opt}$ of the first bond population of a slip-slip bond cluster, such that 
the maximum stability enhancement $\chi_\mathrm{max}$ lies exactly at $\gratio$, can be found from the equality $\fc_1=\fc_2$ as
\begin{align}
\label{ropt}
&\rho\Nt\,\textrm{pln}(\gamma^\prime_1/e) = (1-\rho)\Nt\,\textrm{pln}(\gamma^\prime_2/e) \nonumber \\ \Rightarrow 
&\rho_\mathrm{opt} = \frac{\textrm{pln}(\gamma^\prime_2/e)}{\textrm{pln}(\gamma^\prime_1/e)+\textrm{pln}(\gamma^\prime_2/e)}.
\end{align}
Here, we can also compute $\gratio_\mathrm{opt}$ corresponding to $\chi_\mathrm{max}$ for a fixed 
$\rho$. 

The value of $\chi_\mathrm{max}$ for a slip-slip bond cluster can be found analytically as follows. 
For simplicity, we assume that $\kzero_1 = \kzero_2 = \kzero$, so that $\gamma^\prime_1 = \gamma_1$ and $\gamma^\prime_2 = \gamma_2$. 
First of all, $\chi_\mathrm{max}$ lies at the $\kr_\textrm{opt}$ line, as $\kr_\textrm{opt}$ represents $\kr$ values 
that correspond to the maximum $\fc$ for fixed $\gratio$ and $\fm$ is independent of $\kr$. Therefore, 
we restrict further analysis of $\chi$ to the $\kr_\textrm{opt}$ line, on which $f_1 = \fc_1$, $N_1=\Nc_1$, 
$f_2 = \fc_2$, and $N_2=\Nc_2$.   

For $\gratio < \gratio_\mathrm{opt}$, $\fc_1 < \fc_2$ which implies that 
\begin{align}
\label{chi_1}
\chi \left( \gratio < \gratio_\mathrm{opt} \right) &= \frac{\fc}{\fm} = \frac{\fc}{\fc_2} = \frac{\Nc_1\kr_\textrm{opt} + \Nc_2}{\Nc_2} \nonumber \\&= 
1 + \frac{\rho}{1-\rho}\,\frac{\textrm{pln}(\gamma_1/e)}{\textrm{pln}(\gamma_2/e)}. 
\end{align}
Note that $\chi \left( \gratio < \gratio_\mathrm{opt} \right)$ is a monotonically increasing function of 
$\gamma_1 = \gratio \gamma_2$ or $\gratio$. For $\gratio > \gratio_\mathrm{opt}$, $\fc_1 > \fc_2$, leading to 
\begin{align}
\label{chi_2}
\chi \left( \gratio > \gratio_\mathrm{opt}\right) &= \frac{\fc}{\fm} = \frac{\fc}{\fc_1} = 
\frac{\Nc_1\kr_\textrm{opt} + \Nc_2}{\Nc_1 \kr_\textrm{opt}} \nonumber \\&= 
1 + \frac{1-\rho}{\rho}\,\frac{\textrm{pln}(\gamma_2/e)}{\textrm{pln}(\gamma_1/e)}. 
\end{align}   
$\chi \left( \gratio > \gratio_\mathrm{opt} \right)$ is a monotonically decreasing function of $\gratio$. The monotonic behavior of $\chi$ along 
the $\kr_\textrm{opt}$ line for these two cases of $\gratio$ proves that $\chi_\mathrm{max}$ is achieved exactly at 
the $\gratio_\mathrm{opt}$ value, where $\fc_1=\fc_2$. Then, if we plug the expression for $\rho_\mathrm{opt}$ from Eq.~(\ref{ropt}) 
into Eq.~(\ref{chi_1}) or (\ref{chi_2}), we obtain that 
\begin{equation}
\chi_\mathrm{max} = 2 
\end{equation}
for a slip-slip bond cluster with two bond populations.

To generalize the result for a slip-slip bond cluster with two distinct populations, a cluster with three different slip bond populations is considered. 
For simplicity, we assume that $\kzero_1 = \kzero_2 = \kzero_3 = \kzero$, so that $\gamma^\prime_1 = \gamma_1$,  $\gamma^\prime_2 = \gamma_2$, 
and $\gamma^\prime_3 = \gamma_3$. In the mean-field description, the average number of bonds is determined by
\begin{equation}
\dfrac{\der N_1}{\der\tau} = -N_1\,e^{f_1/(N_1 \fd_1)}+(\Nt_1-N_1)\gamma_1,
\end{equation}
\begin{equation}
\dfrac{\der N_2}{\der\tau} = -N_2\,e^{f_2/(N_2 \fd_2)}+(\Nt_2-N_2)\gamma_2, 
\end{equation}
\begin{equation}
\dfrac{\der N_3}{\der\tau} = -N_3\,e^{f_3/(N_3 \fd_3)}+(\Nt_3-N_3)\gamma_3, 
\end{equation}
where $\Nt_1+\Nt_2+\Nt_3 = \Nt$. Similarly to the case with two sub-clusters, the force balance results in
\begin{equation}
f_1 = \frac{N_1 \kr_1 F}{N_1\kr_1 + N_2\kr_2+N_3},
\end{equation}
\begin{equation}
f_2 = \frac{N_2 \kr_2 F}{N_1\kr_1 + N_2\kr_2+N_3},
\end{equation}
\begin{equation}
f_3 = \frac{N_3 F}{N_1\kr_1 + N_2\kr_2+N_3},
\end{equation}
where $\kr_1 = k_1/k_3$ and $\kr_2 = k_2/k_3$. Similarly, we define $\gratio_1=\gamma_1/\gamma_3$ and $\gratio_2=\gamma_2/\gamma_3$.
Then, the point where $\fc_1=\fc_2=\fc_3 = f_1 = f_2 = f_3$ is considered, as it corresponds to $\chi_\mathrm{max}$, which can be 
shown by arguments similar to those for a cluster with two bond populations. Fractions of the three populations are 
$1-\rho_2-\rho_3$, $\rho_2$, and $\rho_3$, where $\rho_2 + \rho_3 \leq 1$. 

Then, $\chi_\mathrm{max} = \fc / \mathrm{max}(\fc_1,\fc_2,\fc_3) = \fc / \fc_3$ is calculated as
\begin{align}
\label{chi_m3}
\chi_\mathrm{max} &= 1 + \frac{\Nc_1{\kr_\textrm{opt}}_1}{\Nc_3}+
\frac{\Nc_2{\kr_\textrm{opt}}_2}{\Nc_3}  \\ &= 
1+ \frac{1-{\rho_\mathrm{opt}}_2-{\rho_\mathrm{opt}}_3}{{\rho_\mathrm{opt}}_3} 
\, \frac{\mathrm{pln}(\gamma_1/e)}{\mathrm{pln}(\gamma_3/e)} + \frac{{\rho_\mathrm{opt}}_2}{{\rho_\mathrm{opt}}_3} 
\, \frac{\mathrm{pln}(\gamma_2/e)}{\mathrm{pln}(\gamma_3/e)}, \nonumber
\end{align}
where 
\begin{equation}
{\kr_\textrm{opt}}_1 = \frac{1+\textrm{pln}(\gamma_1/e)}{1+\textrm{pln}(\gamma_3/e)}, \,\,\,\,\,\,\,\,\,\, \nonumber
{\kr_\textrm{opt}}_2 = \frac{1+\textrm{pln}(\gamma_2/e)}{1+\textrm{pln}(\gamma_3/e)},  \nonumber
\end{equation}
\begin{equation}
{\rho_\mathrm{opt}}_2 = \frac{\textrm{pln}(\gamma_1/e) \textrm{pln}(\gamma_3/e)}{f(\gamma_1,\gamma_2,\gamma_3)}, \nonumber
\end{equation}
\begin{equation}
{\rho_\mathrm{opt}}_3 = \frac{\textrm{pln}(\gamma_1/e) \textrm{pln}(\gamma_2/e)}{f(\gamma_1,\gamma_2,\gamma_3)}, \nonumber
\end{equation}
and $f(\gamma_1,\gamma_2,\gamma_3) = \textrm{pln}(\gamma_1/e)  \textrm{pln}(\gamma_2/e) +  \textrm{pln}(\gamma_1/e) 
\textrm{pln}(\gamma_3/e)  + \textrm{pln}(\gamma_2/e) \textrm{pln}(\gamma_3/e)$.
By plugging ${\rho_\mathrm{opt}}_2$ and ${\rho_\mathrm{opt}}_3$ into Eq.~(\ref{chi_m3}), we obtain that
\begin{equation}
\chi_\mathrm{max} = 3 
\end{equation} 
for a slip-slip bond cluster with three different bond populations. This argument can be extended to $m$ different 
sub-populations sharing the same load, whose maximum stability enhancement is equal to $m$.


%

\end{document}